\documentclass[prb,superscriptaddress,floatfix,a4paper,aps,twocolumn,nofootinbib]{revtex4-2}

\usepackage{graphicx}
\usepackage{amsmath}
\usepackage{amssymb}
\usepackage[usenames, dvipsnames]{xcolor}
\usepackage[colorlinks=true,citecolor=blue,linkcolor=blue,urlcolor=blue]{hyperref}

\usepackage{tabularx}
\usepackage{color}
\usepackage{xspace}
\usepackage{hyperref}

\begin{document}

\title{Superconducting density of states and bandstructure at the surface of the candidate topological superconductor Au$_2$Pb}

\author{Francisco Mart\'in-Vega}
\affiliation{Laboratorio de Bajas Temperaturas y Altos Campos Magn\'eticos, Departamento de F\'isica de la Materia Condensada, Instituto Nicol\'as Cabrera and Condensed Matter Physics Center (IFIMAC), Unidad Asociada UAM-CSIC, Universidad Aut\'onoma de Madrid, E-28049 Madrid,
Spain}

\author{Edwin Herrera}
\affiliation{Laboratorio de Bajas Temperaturas y Altos Campos Magn\'eticos, Departamento de F\'isica de la Materia Condensada, Instituto Nicol\'as Cabrera and Condensed Matter Physics Center (IFIMAC), Unidad Asociada UAM-CSIC, Universidad Aut\'onoma de Madrid, E-28049 Madrid,
Spain}

\author{Beilun Wu}
\affiliation{Laboratorio de Bajas Temperaturas y Altos Campos Magn\'eticos, Departamento de F\'isica de la Materia Condensada, Instituto Nicol\'as Cabrera and Condensed Matter Physics Center (IFIMAC), Unidad Asociada UAM-CSIC, Universidad Aut\'onoma de Madrid, E-28049 Madrid,
Spain}

\author{V\'ictor Barrena}
\affiliation{Laboratorio de Bajas Temperaturas y Altos Campos Magn\'eticos, Departamento de F\'isica de la Materia Condensada, Instituto Nicol\'as Cabrera and Condensed Matter Physics Center (IFIMAC), Unidad Asociada UAM-CSIC, Universidad Aut\'onoma de Madrid, E-28049 Madrid,
Spain}

\author{Federico Mompe\'an}
\affiliation{Instituto de Ciencia de Materiales de Madrid, Consejo Superior de
Investigaciones Cient\'{\i}ficas (ICMM-CSIC), Unidad Asociada UAM-CSIC, Sor Juana In\'es de la Cruz 3,
E-28049 Madrid, Spain}

\author{Mar Garc{\'i}a-Hern{\'a}ndez}
\affiliation{Instituto de Ciencia de Materiales de Madrid, Consejo Superior de
Investigaciones Cient\'{\i}ficas (ICMM-CSIC), Unidad Asociada UAM-CSIC, Sor Juana In\'es de la Cruz 3,
E-28049 Madrid, Spain}

\author{Paul C. Canfield}
\affiliation{Ames Laboratory, Ames and Department of Physics $\&$ Astronomy, Iowa State University, Ames, IA 50011}

\author{Annica M. Black-Schaffer}
\affiliation{Department of Physics and Astronomy, Uppsala University, Box 516, S-751 20 Uppsala, Sweden}

\author{Jos\'e J. Baldov\'i}
\affiliation{Instituto de Ciencia Molecular (ICMol), Universidad de Valencia, Catedr\'atico Jos\'e Beltr\'an 2, 46980 Paterna, Spain}
\affiliation{Max Planck Institute for the Structure and Dynamics of Matter, Luruper Chaussee 149, D-22761 Hamburg, Germany}

\author{Isabel Guillam\'on}
\affiliation{Laboratorio de Bajas Temperaturas y Altos Campos Magn\'eticos, Departamento de F\'isica de la Materia Condensada, Instituto Nicol\'as Cabrera and Condensed Matter Physics Center (IFIMAC), Unidad Asociada UAM-CSIC, Universidad Aut\'onoma de Madrid, E-28049 Madrid,
Spain}

\author{Hermann Suderow}
\affiliation{Laboratorio de Bajas Temperaturas y Altos Campos Magn\'eticos, Departamento de F\'isica de la Materia Condensada, Instituto Nicol\'as Cabrera and Condensed Matter Physics Center (IFIMAC), Unidad Asociada UAM-CSIC, Universidad Aut\'onoma de Madrid, E-28049 Madrid,
Spain}

\begin{abstract}
The electronic bandstructure of Au$_2$Pb has a Dirac cone which gaps when undergoing a structural transition into a low temperature superconducting phase. This suggests that the superconducting phase ($T_c=1.1$ K) might hold topological properties at the surface. Here we make Scanning Tunneling Microscopy experiments on the surface of superconducting Au$_2$Pb. We measure the superconducting gap and find a sizeable superconducting density of states at the Fermi level. We discuss possible origins for this finding in terms of superconductivity induced into surface states.
\end{abstract}

\maketitle



\section{Introduction}

Recent efforts in band structure calculations have shown that it might be possible to obtain many intermetallic compounds with topologically non-trivial surface states \cite{Zhang2019,Vergniory2019,Tang2019}. Among these systems, a few become superconducting at low temperatures. It has been shown that superconductivity induced by proximity into a topologically non-trival electronic surface state can lead to triplet correlations and the appearance of Majorana modes at the surface \cite{PhysRevB.84.195442,doi:10.1146/annurev-conmatphys-030212-184337,Alicea_2012,PhysRevLett.100.096407}.

\begin{figure}[htbp]
\includegraphics[width = \columnwidth]{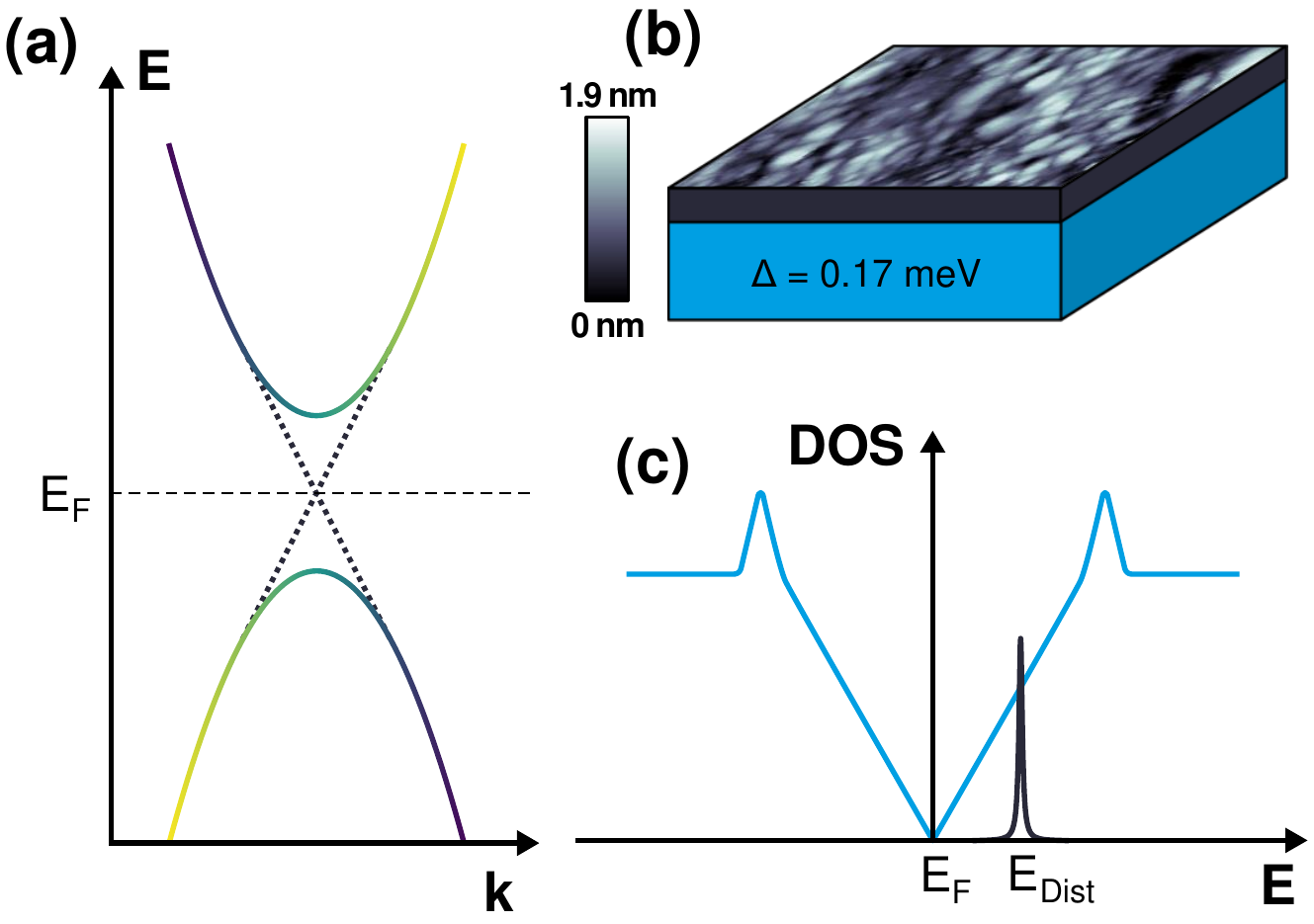}
\caption{(a) Schematic representation of the band inversion occuring in Au$_2$Pb upon cooling into the orthorhombic low temperature phase \cite{Schoop2015,Wu2018}. The color code represents the relation between Pb-sp$^3$ and Au-d character from blue to yellow. At the surface, it is possible that the inverted gap closes (dashed black lines). (b) Schematic representation of our findings. The bulk of Au$_2$Pb is shown in blue and the surface in grey. We find a bulk superconducting gap with $\Delta=0.17$ meV and a surface with a rugosity (white to black spots on the top, following scale bar on the left) which is of the order of a unit cell size and consists of a random height variation. We also find that the surface presents a superconducting density of states with a finite value at zero energy. In (c) we show schematically the superconducting density of states of a system with a surface Dirac cone dispersion inside the superconducting gap (blue) and the localized states generated by scattering (dark grey) at a certain energy $E_{Dist}$ in the vicinity of the Fermi level $E_F$. In a system with a large coherence length and a random distribution of localized states at many different $E_{Dist}$, these smear the low energy part of the density of states into a gapless regime\cite{doi:10.1080/00018732.2014.927109}.
}
\label{Schematics}
\end{figure}

Here we analyze the compound Au$_2$Pb. Au$_2$Pb is a binary intermetallic system which crystallizes at room temperature in a cubic Laves phase \cite{Schoop2015,PhysRevB.93.045118}. At low temperatures, Au$_2$Pb presents an orthorhombic phase which is superconducting with $T_c=1.1$ K \cite{HAMILTON1965665}. Band structure calculations of the room temperature cubic phase predict a bulk 3D Dirac cone due to symmetry allowed band crossings \cite{Schoop2015,PhysRevB.93.045118}. The presence of the bulk Dirac cone is confirmed by angle resolved photoemission studies (ARPES) \cite{Wu2018}. The lowered crystal symmetry of the low temperature phase lifts the band degeneracy and a gap opens at the bulk Dirac cone \cite{Schoop2015,Wu2018}. As we show in detail below and represent schematically in Fig.\,\ref{Schematics}(a), the parity of the bands is inverted at the gap, with the valence band having Au d-electron character and the conduction band Pb-sp$^3$ character.

Experiments aiming to characterize surface states in the normal phase using optical conductivity have concluded that the contribution of the Dirac cone to the optical properties is masked by bulk bands crossing the Fermi energy \cite{Schoop2015,Wu2018,Kemmler_2018}. In the superconducting phase, low temperature specific heat and thermal conductivity measurements have suggested that the bulk superconducting state of Au$_2$Pb is fully gapped, agreeing with s-wave superconductivity \cite{Schoop2015, Yu2016}. Here we analyze the superconducting properties using Scanning Tunnelling Microscope (STM) experiments on the surface of Au$_2$Pb. We measure the superconducting density of states as a function of temperature and magnetic field. We find a superconducting gap value corresponding to a $T_c=1.1$ K within BCS theory and rough surfaces (Fig.\,\ref{Schematics}(b)). The zero bias tunneling conductance is larger than about a third of the normal state density of states. We analyze the possible origin for the finite zero bias tunneling conductance in terms of the influence of defects on surface superconducting states, creating in-gap states schematically represented Fig.\,\ref{Schematics}(c).

\section{Experiments and methods}

\begin{figure}[htbp]
\includegraphics[width =0.8\columnwidth]{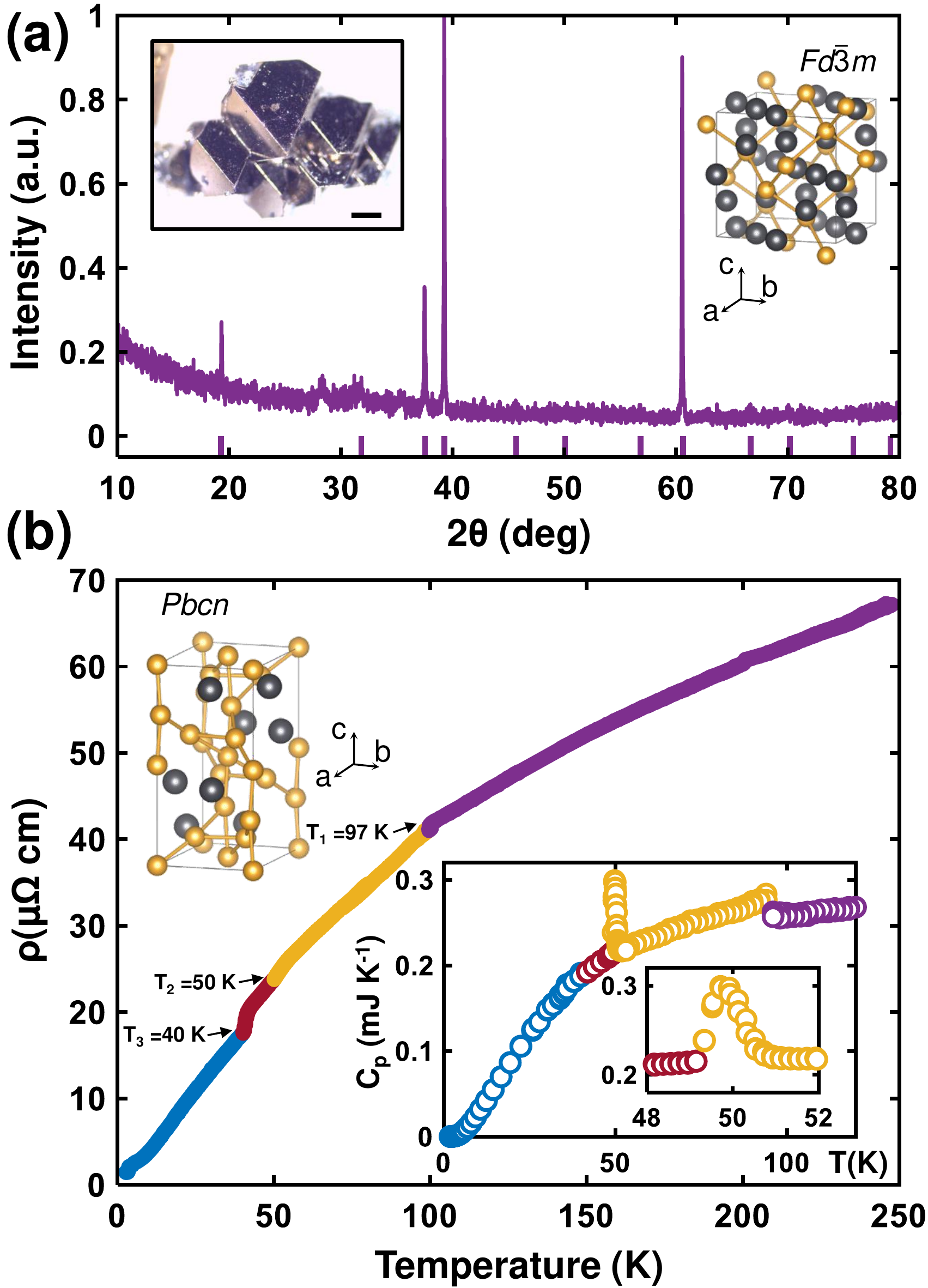}
\caption{(a) Powder x-ray data at room temperature of Au$_2$Pb (purple lines). We show the expected Bragg peaks by vertical lines on the bottom. A picture of the crystals is shown in the upper right inset. We also show a picture of atomic arrangements (Au is shown in gold color and Pb in grey) of the room temperature cubic phase. (b) Resistivity vs temperature of Au$_2$Pb is shown as colored points. The specific heat vs temperature is shown in the inset (with a zoom for the transition at 50 K as an additional inset). We use different colors for the different structural phases. The low temperature crystalline structure (points in blue) is shown in the upper left inset.}
\label{FigSample}
\end{figure}

We synthesized the sample with a Pb flux growth method \cite{doi:10.1080/13642819208215073,Canfield_2019,Schoop2015,Wu2018}. Lumps of gold (99.99\% Sempsa) and lead (99.999\% Goodfellow) with a molar ratio Au:Pb 42:58 were put inside frit disk alumina crucibles \cite{doi:10.1080/14786435.2015.1122248} and sealed in an evacuated silica ampoule. The ampoule was then heated in 12 hours to 1150$^{\circ}$C, fast-cooled to 375$^{\circ}$C in 12 hours, held at this temperature for 6 hours and then slow-cooled to 275$^{\circ}$C in 22 hours. At this point the excess flux was decanted using a centrifuge. The obtained crystals typically had sizes of a few millimeters and showed clean triangular dark-silvered facets. We show powder x-ray data in Fig.\,\ref{FigSample}(a) and a picture of the samples in the upper left inset. In Fig.\,\ref{FigSample}(b) we show the resistivity as a function of temperature, and highlight with colors the different crystalline structures\cite{PhysRevB.93.045118,Schoop2015}. In the lower right inset of Fig.\,\ref{FigSample}(b) we show the specific heat vs temperature. These data mostly coincide with previous results in high quality single crystals of Au$_2$Pb\cite{Schoop2015,Wu2018}. Furthermore, in the insets of Fig.\,\ref{FigSample}(a,b) we show the room and low temperature crystalline structures found in Au$_2$Pb. At room temperature (violet points in Fig.\,\ref{FigSample}(b)), the crystal structure is the cubic Laves phase $Fd\overline{3}m$ (no. 227) \cite{PhysRevB.93.045118}. It is based on the pyrochlore lattice, with Au atoms having a tetrahedric arrangement distributed inside a face centered cubic lattice of Pb atoms. The crystalline structures of the phase at intermediate temperatures (yellow and red points in Fig.\,\ref{FigSample}(b)) are unknown. The structure of the low temperature phase (blue points in Fig.\,\ref{FigSample}(b)) is derived from the cubic room temperature phase by small atomic distortions, giving orthorhombic $Pbcn$, No. 60 \cite{Schoop2015}.

We performed STM measurements using a home-made microscope described in Ref.\,\onlinecite{doi:10.1063/1.3567008} and the software described in Ref.\,\cite{doi:10.1063/5.0064511}. The energy resolution of the tunneling conductance measurement setup is of 10 $\mu$eV as shown measurements in Al with a similar set-up ($T_c=1.1$ K) \cite{doi:10.1063/5.0059394}. For image treatment, we use Refs.\,\cite{doi:10.1063/5.0064511,doi:10.1063/1.2432410}. We cut single crystalline samples in rectangular bars that are a few mm long and a fraction of a mm wide and mount them on the STM sample holder. The position of the sample holder can be modified to approach the long axis of the bar shaped sample to a blade in-situ at 4 K in such a way that the blade cleaves the sample and exposes a fresh surface. We exposed a surface perpendicular to a main crystalline axis of the cubic room temperature crystalline structure. Given the different structural transitions undergone upon cooling we can expect that we maintain a crystalline face at low temperatures. We observed large and flat areas over fields of view of a few hundreds of nm. We did not observe atomic resolution, but instead surfaces with elongated structures and a surface corrugation of order of the unit cell size (between 1 nm and 2 nm, see Fig.\,\ref{FigHisto}(a) and the Appendix).

First-principles calculations with spin-orbit coupling (SOC) were performed using the Quantum ESPRESSO package \cite{Giannozzi_2009}. In our calculations we employed the generalized gradient approximation (GGA) with the Perdew-Burke-Ernzerhof (PBE) \cite{PhysRevLett.77.3865} exchange-correlation functional. We used fully-relativistic norm-conserving Pseudo DoJo pseudopotentials \cite{VANSETTEN201839} and the electronic wave functions were expanded with well-converged kinetic energy cutoffs of 45 Ry and 180 Ry for the wavefunctions and charge density, respectively. The Brillouin zone was sampled with a Monkhorst-Pack \cite{PhysRevB.13.5188} ($8\times 8 \times 8$) $k$-point mesh. Dispersion interactions to account for van der Waals interactions between the layers were considered by applying semi-empirical Grimme DFT-D3 corrections \cite{https://doi.org/10.1002/jcc.20495}. Structures were fully optimized using the Broyden-Fletcher-Goldfarb-Shanno (BFGS) algorithm \cite{doi:10.1063/1.462844} until the forces on each atom were smaller than $2\cdot 10^4$ Ry/au and the energy difference between two consecutive relaxation steps less than $10^4$ Ry. To analyze the surface bandstructure, we constructed a tight-binding model based on maximally localized Wannier functions \cite{PhysRevB.56.12847,PhysRevB.65.035109,RevModPhys.84.1419} with Au $sd$ and Pb $sp$ orbitals in the range $E_F \pm 1$ eV. We verified that we find the same results as in the first principles calculations within this energy range. To find the structure of surface states, we built surface spectral functions for a semi-infinite Au$_2$Pb(001) surface via the surface Green’s function method \cite{PhysRevB.23.4988,PhysRevB.23.4997,Sancho_1984,Sancho_1985} as implemented in the WannierTools package \cite{WU2018405}.

\section{Results}

\begin{figure}[htbp]
\includegraphics[width = \columnwidth]{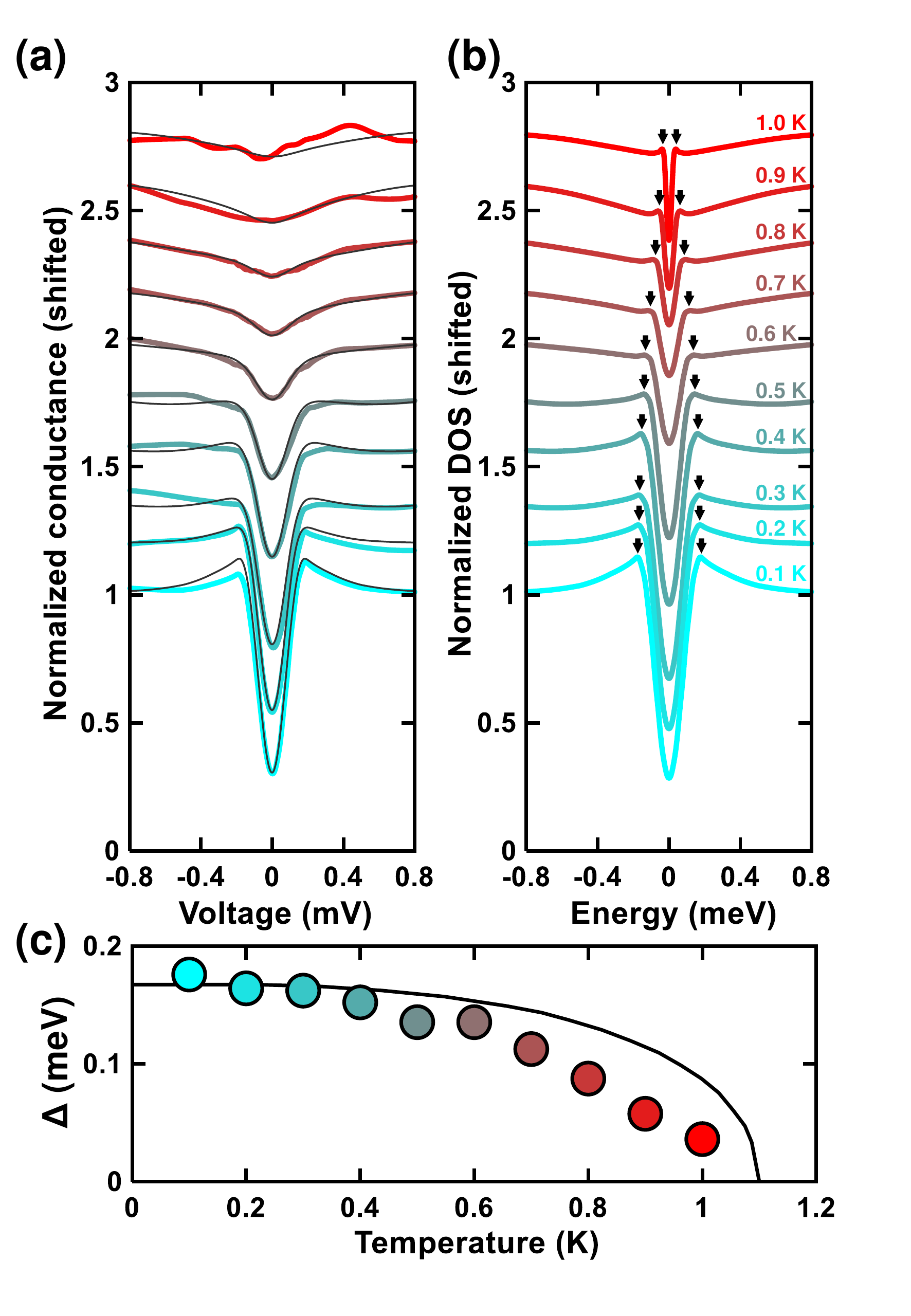}
\caption{(a) Tunneling conductance as a function of the bias voltage for different temperatures (colored lines, bottom is 100 mK, in light blue, and top is 1 K, in red, with 100 mK increases). The tunneling conductance is normalized to the value at bias voltages between 0.5 mV and 1 mV. (b) Local density of states (LDOS) as a function of energy for the same temperatures and using the same color scheme. Black lines in (a) provide the LDOS shown in (b) convoluted with temperature. (c) Temperature dependence of the superconducting gap, also shown by the arrows in (b) The BCS temperature dependence of the superconducting gap is shown by a black line.
}
\label{FigZeroField}
\end{figure}

\begin{figure}[htbp]
\includegraphics[width = \columnwidth]{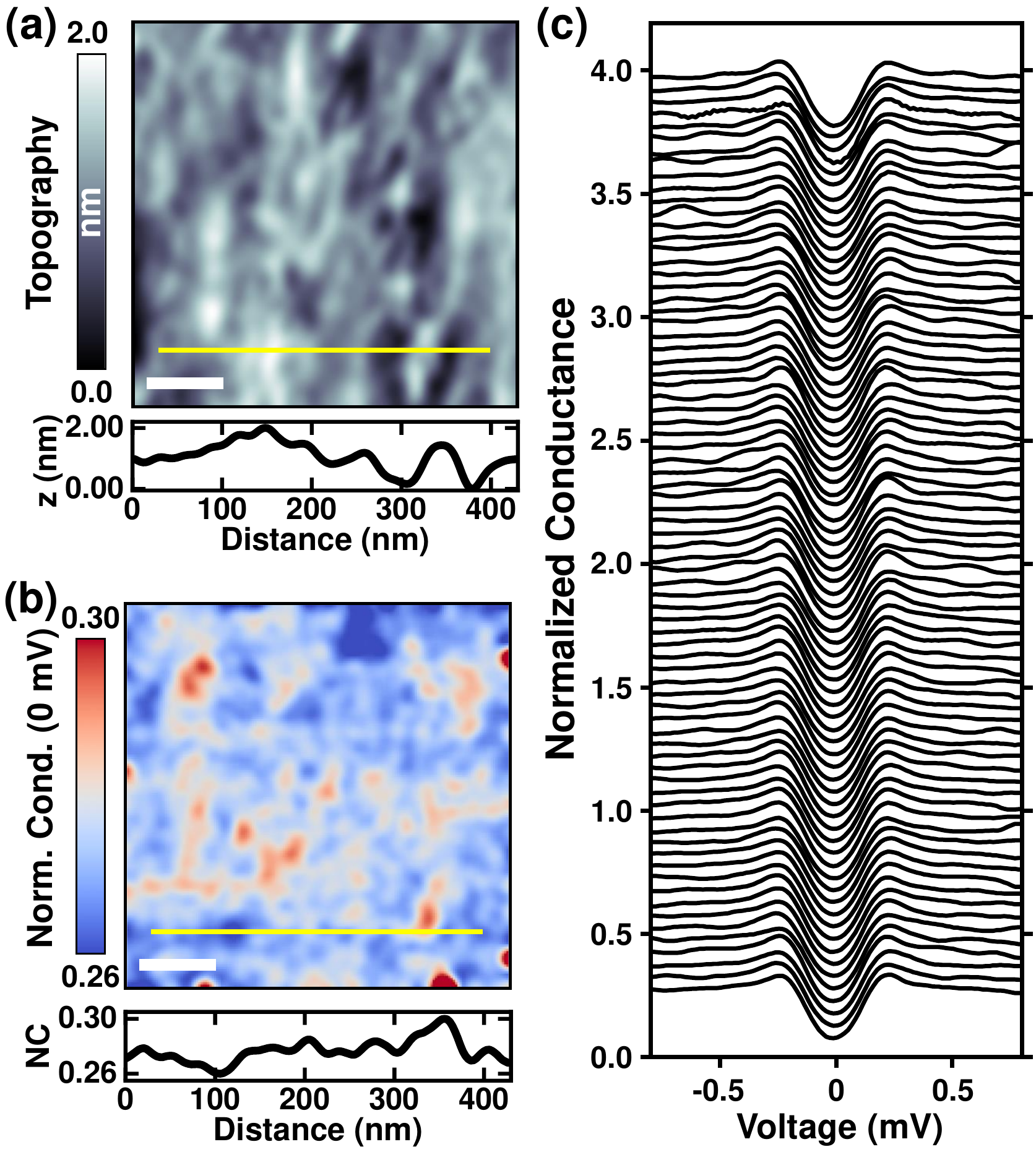}
\caption{(a) Scanning Tunneling Microscopy image obtained with a constant current of 1 nA and at a bias voltage of 10 mV. The obtained height changes are shown as a grey scale, following the bar on the left. The height profile along the yellow line is shown in the bottom panel. (b) Tunneling conductance map at zero bias in the same field of view as in (a). Conductance changes are shown as a blue-red variation following the bar on the left. The conductance profile along the yellow line is shown in the bottom panel. White scale bars in (a,b) are 60 nm long. (c) Tunneling spectroscopy as a function of the bias voltage along the yellow lines in (a,b).
}
\label{FigHisto}
\end{figure}

\begin{figure*}[htbp]
\includegraphics[width = \textwidth]{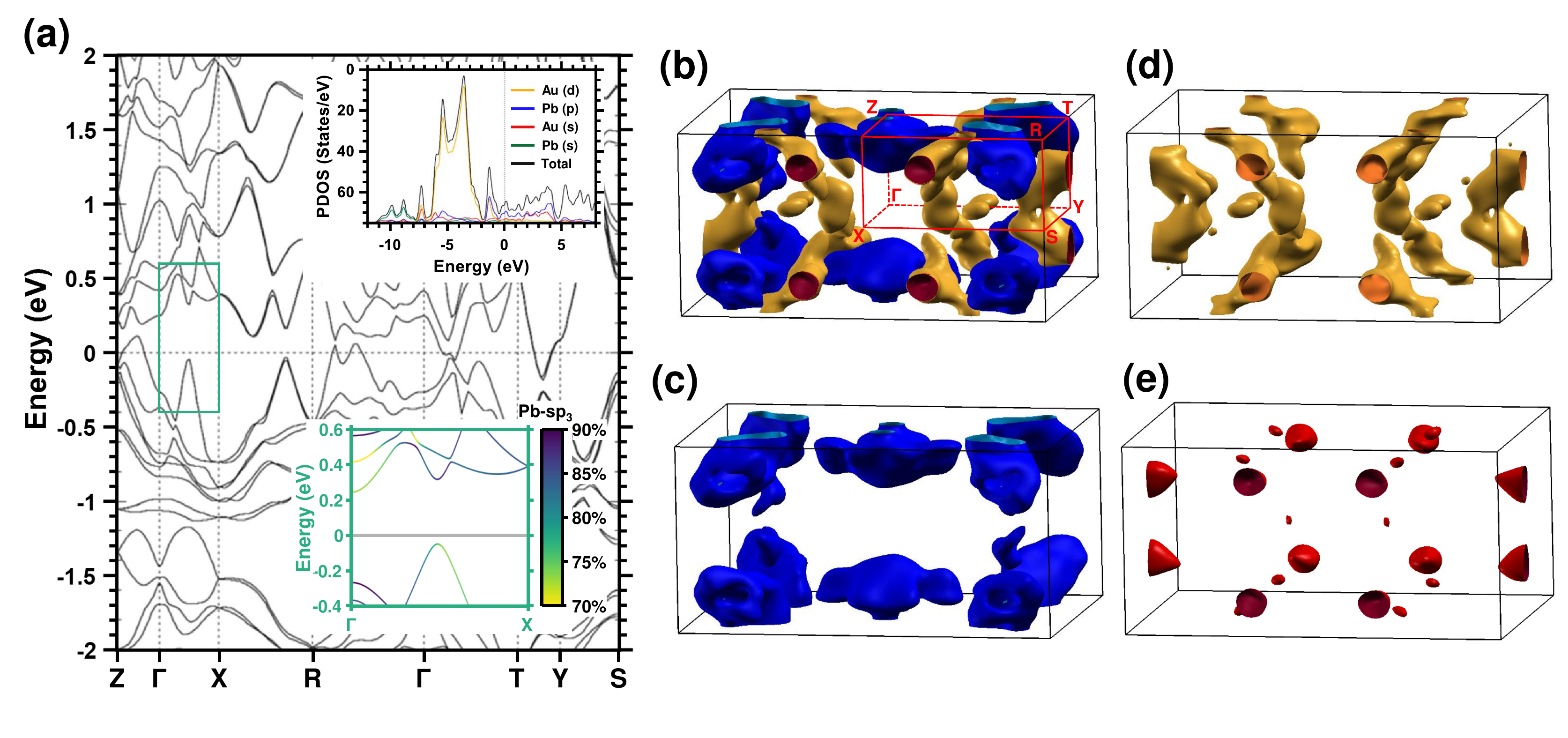}
\caption{(a) The electronic bandstructure of the low temperature orthorhombic phase of Au$_2$Pb, obtained from density functional calculations is shown by the black lines. In the upper right inset we show as lines the partial densities of states of the orbitals marked in the legend as a function of the energy. We see that, at the Fermi level, the largest contributions are due to Pb (p), Au (s) and Au (d) orbitals. In the lower left inset we show a zoom over the $\Gamma$-$X$ direction, with the gap that opens in the low temperature phase and the associated band inversion \cite{Schoop2015,Wu2018}. The blue to yellow color scale shows the relation between Pb-sp$_3$ and Au d-orbital character, following the scale bar at the right. (b) Fermi surface of the orthorhombic low temperature phase of Au$_2$Pb is shown as colored surfaces. The first Brillouin zone, with the high symmetry directions, is shown in red. The Fermi surface is made by three bands, the 15$^{th}$ (in blue, (c)), the 17th$^{th}$ (in yellow, (d)) and the 19th$^{th}$ band (in red, (e)).}
\label{FigTeo1}
\end{figure*}

In Fig.\,\ref{FigZeroField} we show Scanning Tunneling Spectroscopy measurements of the superconducting gap of Au$_2$Pb as a function of temperature. We see that there is no fully opened superconducting gap, but instead a large zero bias conductance. The quasiparticle peaks are also supressed with respect to the expected high peaks for a usual s-wave BCS superconductor. We deconvolute the local density of states ($N(E)$) at each temperature, using the relation $G(V)=\int N(E)\frac{df(E-eV)}{dV} dE$ (see Refs.\,\cite{PhysRevLett.96.027003,PhysRevLett.101.166407,PhysRevB.97.134501,doi:10.1063/5.0064511} and the Appendix) and show the result in Fig.\,\ref{FigZeroField}(b). From the density of states, we obtain the superconducting gap $\Delta$ and its temperature dependence by following the quasiparticle peak positions in $N(E)$. We find a superconducting gap of $\Delta(T=0K)=0.17$ meV. The BCS expression $\Delta(T=0$ K$)=1.76k_BT_c$ provides $\Delta(T=0$ K$)\approx 0.17$meV using $T_c=1.1$ K, in close agreement with the value we find. Fig.\,\ref{FigZeroField}(c) shows the temperature dependence of the superconducting gap $\Delta(T)$ up to $T_c=1.1$ K. $\Delta(T)$ roughly follows the temperature dependence of a fully opened s-wave superconducting gap expected within BCS theory.

The variations of the zero bias conductance as a function of the position are shown in Fig.\,\ref{FigHisto}(b). We observe variations that occur randomly over the whole surface. These are small, below about 10\% of the normal state tunneling conductance. The magnetic field dependence is shown in the Appendix and is compatible with superconductivity having a low Ginzburg-Landau parameter $\kappa$ and a coherence length of about $\xi\approx$ 60 nm, as shown previously\cite{Schoop2015,Wu2018}. Thus, we find a gap value which is compatible with specific heat measurements \cite{Schoop2015,Xing2016,Wu2018}, but there is a sizeable finite density of states at the Fermi level which slightly varies as a function of the position.

We show the calculated electronic band structure of the low temperature orthorhombic phase in Fig.\,\ref{FigTeo1} (a). The overall shape coincides with previous calculations \cite{Schoop2015}. The Dirac cone of the room temperature cubic structure is opened by a gap along the $\Gamma-X$ line. The Fermi surface presents three bands crossings, the 15$^{th}$, the 17$^{th}$ and the 19$^{th}$ bands. All are spin degenerate and produce three dimensional pockets with an intricate structure (Fig.\,\ref{FigTeo1}(b)). The top and bottom of the bands crossing the Fermi level are just a few hundreds of meV from the Fermi level (Fig.\,\ref{FigTeo1}(a)), confirming that Au$_2$Pb is a low density semimetallic system. The 15$^{th}$ band (Fig.\,\ref{FigTeo1}(c)) consists essentially of two hole pockets located at the top side and the edges of the Brillouin zone. It has a mixed Pb-p and Au-s,d character. The 17$^{th}$ band (Fig.\,\ref{FigTeo1}(d)) consists of three electron pockets. One closed spherical pocket located well inside the Brillouin zone. Another one goes from the top to the bottom of the Brillouin zone, and the third one is located at the sides of the Brillouin zone. The 19$^{th}$ band (Fig.\,\ref{FigTeo1}(e)) produces two electron pockets that are nested inside the 17$^{th}$ band pockets located at the sides of the Brillouin zone. Actually, these two pockets are nearly degenerate, as shown by the bandstructure between the $T$ and $Y$ high symmetry points and both have a predominant Au-s,d character.

\begin{figure}[htbp]
\includegraphics[width = \columnwidth]{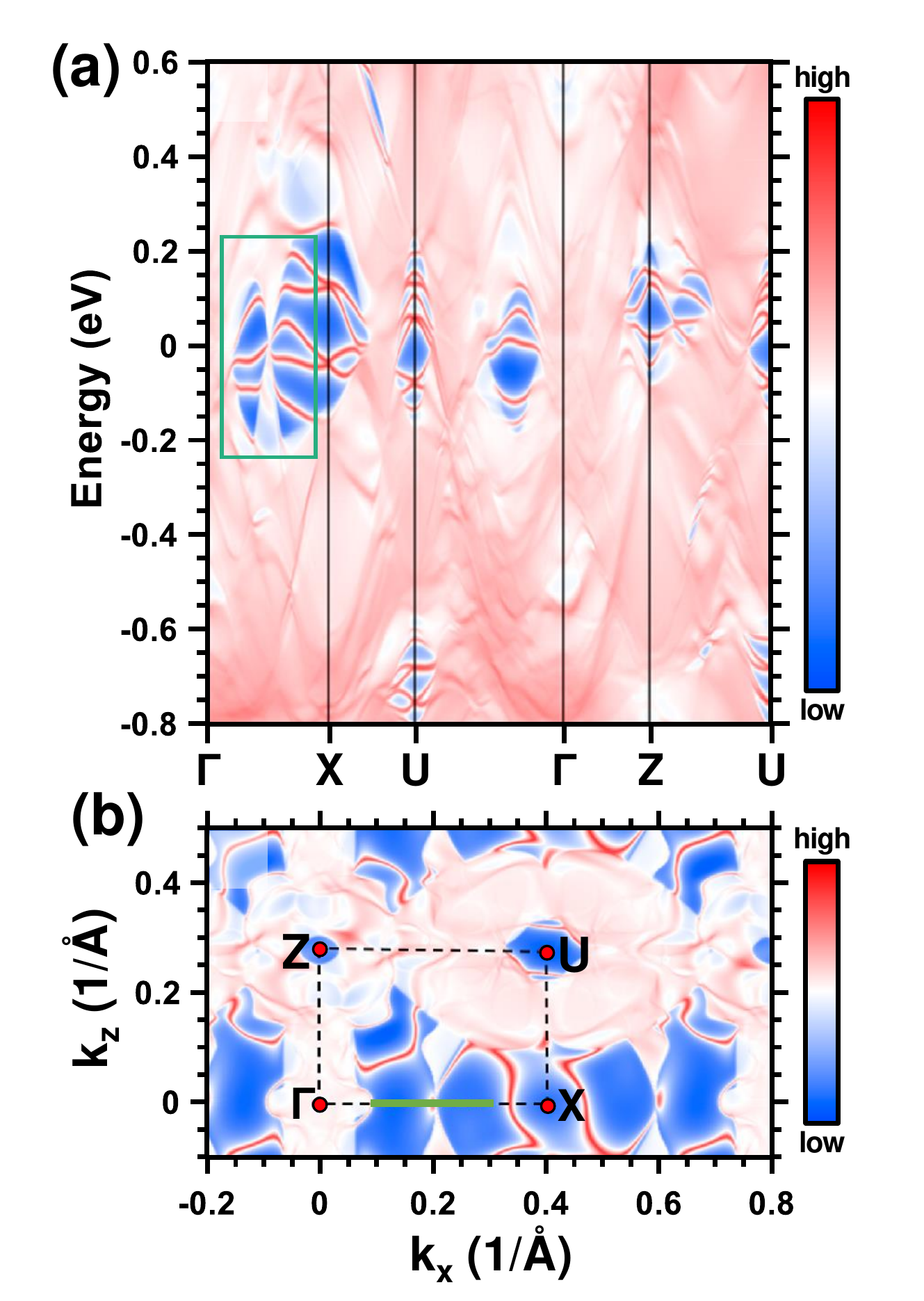}
\caption{Surface bandstructure of the low temperature orthorhombic phase of Au$_2$Pb, with the electronic density shown in blue to red. In (a) we show the result for the (010) surface as a function of energy. In blue are gapped regions without bulk or surface states. The green rectangle shows the region corresponding to the green rectangle in Fig.\,\ref{FigTeo1}(a), highlighted by Refs.\,\cite{Schoop2015,Wu2018}. In (b) we show the Fermi level cut of the surface bandstructure. The Brillouin zone high symmetry directions are shown as red points connected by dashed lines. The green bar shows approximately the region covered by the green rectangle in (a).}
\label{FigTeo2}
\end{figure}

\section{Discussion}

Our tunneling conductance measurements (Fig.\,\ref{FigZeroField}) show a single quasiparticle peak, showing that the gap magnitude is similar over the whole Fermi surface. However, we do not observe the zero density of states expected at a Fermi level for a usual isotropic s-wave superconductor. Furthermore, the measured $LDOS(E)$ at low temperatures (Fig.\,\ref{FigZeroField}(b)) has a continous V-shaped variation for low energies, instead of a U or squarish shaped curve expected for a usual s-wave superconductor. These two findings differ from previous bulk specific heat measurements, which suggest instead a fully opened superconducting gap \cite{Schoop2015, Yu2016}.

One possible origin for modified superconducting properties at the surface is chemical degradation. This can be excluded in our case, because samples were cleaved at low temperatures in fully inert cryogenic vacuum. Thus, the STM measurements show that superconductivity at or close to the surface in Au$_2$Pb is different than in the bulk.

To discuss this issue, we should first note that the surface roughness we find here is of order of the size of a unit cell (c-axis lattice parameter is of 1.1 nm). Previous ARPES experiments have found electronic surface states in polished surfaces of the room temperature cubic phase, whose roughness is above a unit cell size. Thus, the actual surface roughness does not disturb significantly the formation of surface states  \cite{Wu2018}.

As a first scenario to discuss our results, we consider group theory applied to the low temperature orthorhombic phase. The Dirac cone of the high temperature tetragonal phase along the $\Gamma-X$ line opens in the low temperature orthorhomic phase, due to symmetry breaking allowed band mixing. The influence of this situation in superconductivity was recently analyzed in detail for the case of Au$_2$Pb in Ref.\,\cite{Cheon2021}. For the $D_{2h}$ orthorhombic point group (low temperature crystalline phase), only the superconducting states belonging to the trivial $A_g$ and the triplet $A_u$ representations have no nodes in the bulk and are thus compatible with the available bulk specific heat. Out of these, the $A_u$ state presents a linear Dirac band dispersion inside the bulk superconducting gap at the surface. This surface state is topologically protected by a zero-dimensional topological number that can be defined using the $y-z$ mirror symmetry \cite{PhysRevLett.115.187001,PhysRevB.94.014510,Cheon2021}. The linear Dirac band dispersion inside the superconducting gap generates a linearly dependent superconducting density of states, with massless particles and thus zero states at the Fermi level. Atomic scale impurities within such a two-dimensional Dirac spectrum are known to produce distinct impurity resonances at an energy $E_{dist}$ varying with the strength in the impurity and located within the linear portion of the spectrum (see Fig.\,\ref{Schematics}(c)) \cite{doi:10.1080/00018732.2014.927109}. These resonances have been observed inside the Dirac in-gap dispersion of cuprate superconductors and graphene and are inherent to a Dirac dispersion relation \cite{doi:10.1126/science.285.5424.88,PhysRevLett.104.096804,doi:10.1080/00018732.2014.927109}. Thus, one possible origin for a finite zero bias tunneling conductance is that the surface corrugation generates in-gap states at different energies $E_{dist}$ that overlap and provide a finite zero bias tunneling conductance with slight variations as a function of the position. This scenario would imply, however, a triplet bulk superconducting order parameter belonging to the $A_u$ representation. In absence of bulk spin-dependent measurements, such as Knight shift, there is no firm ground to assume triplet superconductivity in Au$_2$Pb.

As a second scenario, we analyze in more detail the surface band structure in the normal state. We have calculated the surface Green's function using DFT on a semi-infinite Au$_2$Pb sample. We find well defined surface states located in the gaps of the bulk bandstructure for all surface terminations. Furthermore, the spin degeneracy of these surface states is lifted by spin orbit coupling. We show results on the (010) surface in Fig.\,\ref{FigTeo2}(a,b). The $\mathbb Z_2$ topological invariant is one on this plane. We can clearly see that the gap along $\Gamma-X$ is closed at the surface by a Dirac cone. In addition, there are three surface states inside the gap, two of which cross the Fermi level at the center of the Dirac cone. These two surface states cross again the Fermi level close to $X$. As we can see in Fig.\,\ref{FigTeo2}(b), there are a number of additional surface states crossing the Fermi level along different directions of the Brillouin zone. Ginzburg showed that it was in principle possible to find superconductivity on a surface two-dimensional electron gas \cite{Ginzburg2D}. It has been shown that surface states of Ag and Au deposited on top of a superconductor show a superconducting density of states with a non-zero value at the Fermi level \cite{PhysRevB.94.220503,PhysRevLett.122.247002}. Our measurements provide a similar result, suggesting that the opening of the superconducting gap at the surface states is incomplete.

We observe in our images that the surface corrugation is well above atomic size. The surface consists of irregular changes in height. However, absence of atomically flat surfaces does not modify surface states, as mentioned above and shown in Ref.\,\cite{Wu2018}. Still, it induces disorder, which in turn smears out any sharp features in the density of states. The density of states we find in our experiments has a strong V-shape that can also be compatible with gapless s-wave superconductivity at surface states, the triplet order parameter discussed above as well as other possibilities that are not compatible with bulk measurements, such as nodal d-wave superconductivity with impurities\cite{doi:10.1080/00018732.2014.927109}. In any case, the density of states we observe is strongly influenced by the disordered surface topography.

\section{Conclusion}

In conclusion, we have determined the superconducting gap and density of states of Au$_2$Pb. We find a superconducting gap magnitude of $0.17$ meV, compatible with $T_c=1.1$ K and BCS relation between the critical temperature and the gap. The superconducting density of states measured at the surface has a V-shape and a large finite value at zero bias, suggesting that there is a finite zero energy density of states. As possible origins for the finite density of states we discuss bulk triplet superconductivity and the concomitant Dirac states at the surface. We also discuss an intricate structure of surface states with incomplete superconducting gap opening.

\section*{Acknowledgements}

This work was supported by the Spanish Research State Agency (PID2020-114071RB-I00, FIS2017-84330-R, CEX2018-000805-M, RYC-2014-15093, MAT2017-87134-C2-2-R), by the Comunidad de Madrid through program NANOFRONTMAG-CM (S2013/MIT-2850), by the European Research Council PNICTEYES grant agreement 679080 and by the Swedish Research Council (Vetenskapsr\r{a}det Grant No. 2018-03488). We acknowledge collaborations through EU program Cost CA16218 (Nanocohybri). J.J.B. acknowledges the Marie Curie Fellowship program (H2020-MSCA-IF2016-751047) and the Generalitat Valenciana (CDEIGENT/2019/022). Ames Laboratory is operated for the U.S. Department of Energy by Iowa State University under Contract No. DE-AC02-07CH11358. We acknowledge SEGAINVEX at UAM for design and construction of cryogenic equipment. We thank Prof. Ángel Rubio for access to computational resources. We thank Marta Sánchez Lomana and Raquel S\'anchez Barquilla for support during crystal growth. We also thank Rafael \'Alvarez Montoya, Sara Delgado and Jos\'e Mar\'ia Castilla for technical support. We acknowledge SIDI at UAM for support in sample characterization.

\newpage

\section*{Appendix}

\begin{figure}[htbp]
\includegraphics[width = \columnwidth]{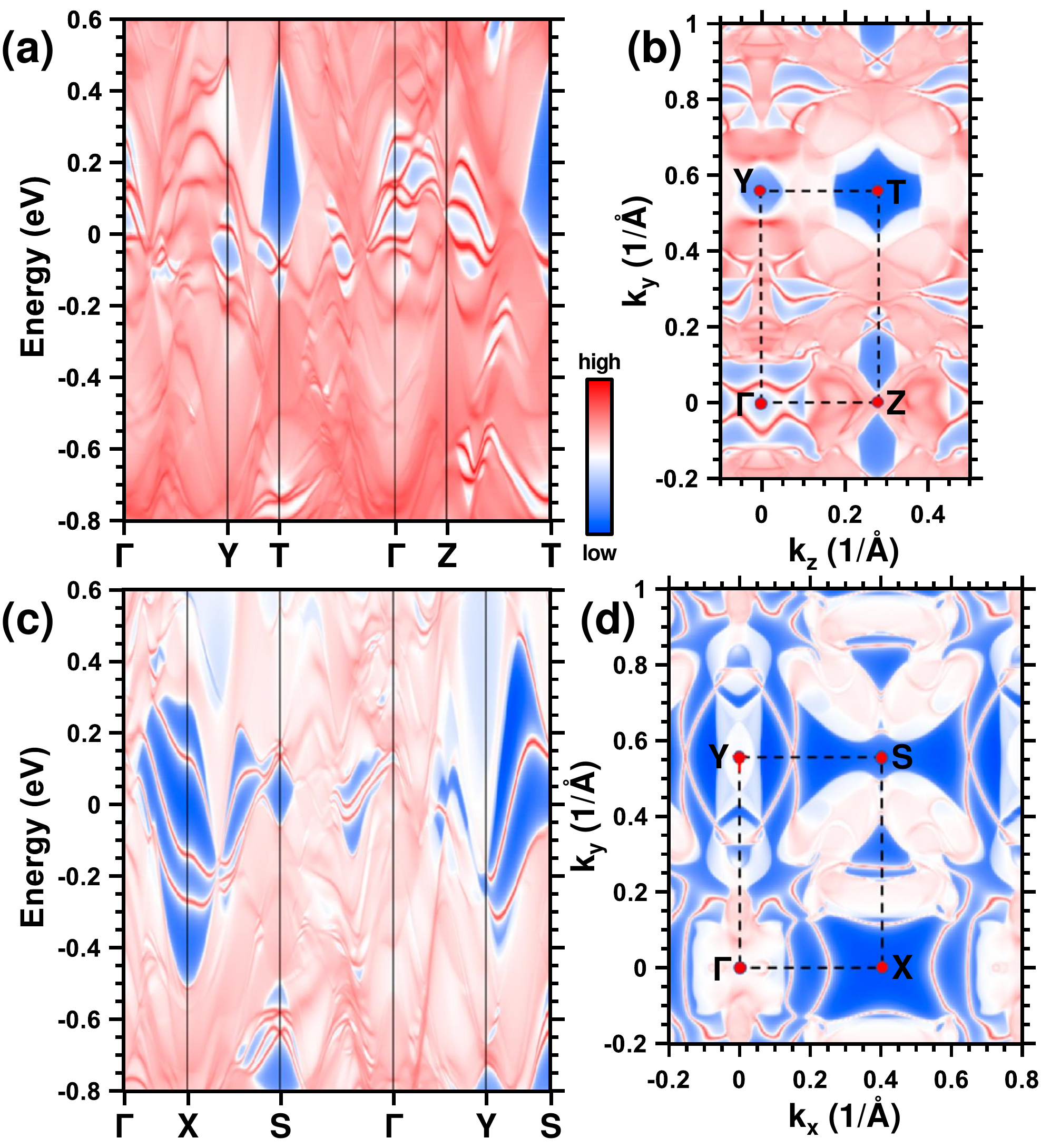}
\caption{In (a) we show the calculated surface bandstructure of the (100) as a function of the energy. The density of states is shown by the red-blue color scale. Surface states appear in bright red color and join areas with projected bulk states. In (b) we show the Fermi energy cut in the Brillouin zone high symmetry points indicated by red dots and dashed lines. The same figure are shown for the (001) surface in (c,d).}
\label{FigTeo3}
\end{figure}

\section*{Calculation of the band structure at the surface}

We show the surface bandstructure at the (100) and (001) surfaces in Fig.\,\ref{FigTeo3}. The gapped region around $U$ remains free of surface states close to the Fermi level. But there are numerous surface states at the Fermi level along all other directions. There is for instance a close surface contour that emerges from the pocket of the 17$^{th}$ band on top and bottom of the Brillouin zone. There is also another state connecting the 15$^{th}$ and 17$^{th}$ bands together along the $\Gamma$-U line and the same occurs along the $\Gamma$-X line.

The $\mathbb Z_2$ topological index of the (100) and (010) surfaces is one, while it is zero for the (001) surface.

\section*{Details on the crystalline structure and the surface topography}

\begin{figure}[htbp]
\includegraphics[width = \columnwidth]{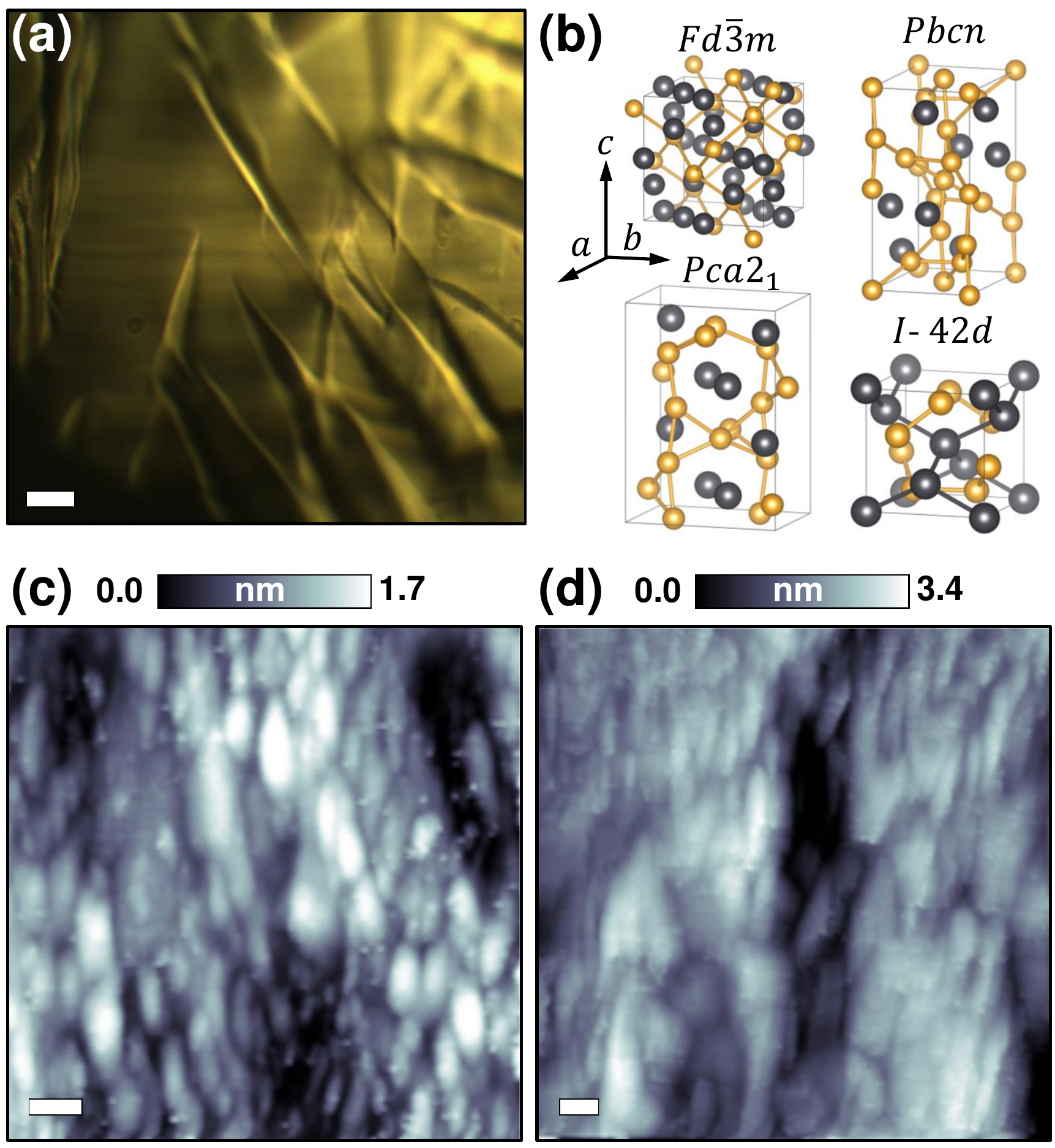}
\caption{(a) Optical picture of the surface of the sample made at room temperature after cleaving the sample at 4 K (scale bar is 10 $\mu$m long). (b) Schematic atomic arrangement, with Au in gold and Pb in grey, in the room temperature structure (cubic laves phase, Fd$\overline{3}$m), the low temperature structure (Pbcn) and two further structures that are found under pressure\cite{Wu2019}. All structures derive from the room temperature cubic laves phase. (c,d) STM images obtained at 100 mK on different fields of view. Scale bars are 20 nm long. The corrugation shown in the STM images is of about 2-3 nm, as shown by the color bars on top of (c,d). 
}
\label{FigSurface}
\end{figure}

In Fig.\,\ref{FigSurface}(a) we present an optical picture of the surface seen after having cleaved the sample at low temperatures. We see triangular like features. We present the room temperature structure of Au$_2$Pb schematically in the left panel of Fig.\,\ref{FigSurface}(b). This is a cubic Laves phase Fd$\overline 3$m with a cell parameter of $a=$ 7.9141(2) \AA \cite{Chen2016}. The structure in the low temperature phase is the Pbcn structure and is derived from the cubic Laves phase by small atomic distortions \cite{Schoop2015}. The intermediate temperature structures (between 40 K and 97 K) are still unknown. Further structural analysis has also lead to the structures Pca2$_1$ and $I\overline{4}2d$ which are also derived from the Laves phase and occur under pressure \cite{Wu2019}. These are shown for completeness in Fig.\,\ref{FigSurface}(b) (bottom panels). Notice that the crystalline transitions induced on cooling are quite varied, as shown in the specific heat (Fig.\,\ref{FigSample}(b)). The transitions at 97 K and at 40 K consist of jumps and the transition at 50 K consists of a large peak, with additional hysteresis, indicating latent heat. The actual atomic arrangements on cooling might lead to changes in the surface crystalline plane. This might influence cleaving, making it difficult to establish a clear cleaving plane at low temperatures, and leading to the absence of atomic size features, as we observe here. The optical picture obtained after cleaving at room temperature (Fig.\,\ref{FigSurface}(a)) shows triangular features, but no clear squarish or linear features as expected in a cubic structure, suggesting that the cleavage was not made along a clear crystalline direction.

We have cleaved two samples and analyzed in each over ten different fields of view, always finding surfaces as discussed in the main text. We often use 10 mV as a bias and work with tunneling currents between 0.1 nA and about 3 nA, finding the same images and tunneling spectroscopy, irrespective of the actual tunneling conductance. We present further images obtained in different fields of view in  Fig.\,\ref{FigSurface}(c,d).

\section*{Tunneling conductance and density of states in a superconductor}

At exactly zero temperature, the tunneling current vs bias voltage obtained with a STM is proportional to the convolution of the densities of states of tip and sample within an energy interval between zero energy and eV, where V is the applied bias voltage and e the elementary charge\cite{Wolf}. As the superconducting gap $\Delta$ is usually much smaller than the Fermi energy $E_F$, $\Delta \ll E_F$, the density of states of the tip is featureless in the energy range of interest in superconducting samples. Thus, the tunneling current is just the integral between zero energy and eV of the density of states of the sample. Derivation leads to the well known relation for the tunneling conductance $G$, $G(V)\propto N_{Superc}(eV)$ where $N_{Superc}(eV)$ is the density of states of the superconductor. In a single gap s-wave superconductor, the density of states is given by $N_{Superc}(eV)=N_{BCS}(E)=\left(\frac{E}{\sqrt{E^2-\Delta^2}}\right)$ for $\vert{E}\vert>\Delta$ and zero elsewhere. At finite temperatures, the Fermi function smears the features of the density of states in the conductance, which is given by $G(V)\propto \int_{-\infty}^{\infty}N_{Superc}(eV)\frac{\partial f(E-eV)}{\partial V}dE$. At the same time, the gap $\Delta$ changes with temperature. For a s-wave BCS superconductor, we can try to follow the main feature of $N_{BCS}(eV)$, the superconducting gap $\Delta(T)$ as a function of temperature. To this end, one might try to trace the maximum in $G(V)$ versus temperature. However, while $\frac{\partial f(E-eV)}{\partial V}$ is a symmetric function at any $eV$, $N_{BCS}(E)$ is definitely not symmetric for $eV\approx \Delta$, because $N_{BCS}(E)\equiv 0$ for $E<\Delta$ and presents a sharp peak for $E>\Delta$. Thus, the maximum in $G(E)$ does not coincide with the maximum in $N_{BCS}(E)$ for finite temperatures $T$. This is valid for features usually found in any $N_{Superc}(E)$ in superconductors that do not follow the simple s-wave single band expression $N_{BCS}(E)$. For example when there are multiple gap peaks in the density of states of multiband superconductors, or for the linearly dispersing $N_{Superc}(E)$ proposed in Fig.\,\ref{Schematics}(c), or for the density of states found in d-wave superconductors. To obtain the temperature dependence of features in $N_{Superc}(E)$, we have to de-convolute $N_{Superc}(eV)$ from $G(V)\propto \int_{-\infty}^{\infty}N_{Superc}(eV)\frac{\partial f(E-eV)}{\partial V}dE$. Using this method, the temperature dependence of the main gap values were found in several multiband superconductors \cite{PhysRevLett.101.166407,PhysRevB.97.134501} and the gap was followed as a function of temperature in a magnetic superconductor with many states inside the gap\cite{PhysRevLett.96.027003}. The method is however best illustrated in a simple s-wave superconductor as Pb. The results in that case are described in Ref.\,\cite{doi:10.1063/5.0064511}.

\begin{figure}[htbp]
\includegraphics[width = \columnwidth]{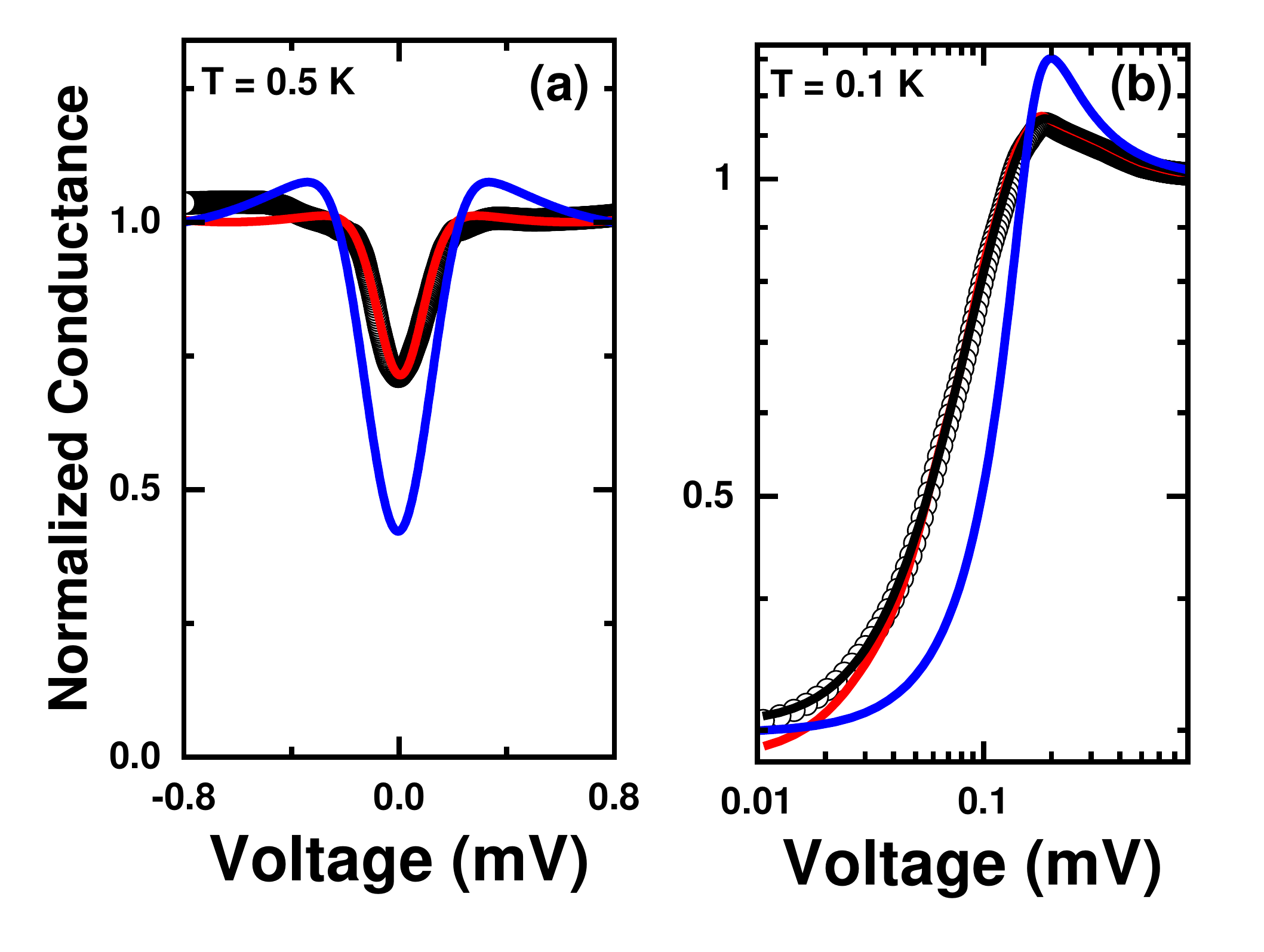}
\caption{(a) The normalized tunneling conductance obtained in the experiment at 0.5 K is shown as black disks. The conductance obtained from the convolution of $N_{Superc}(E)$, shown in Fig.\,\ref{FigZeroField}(b) at 0.5 K, and the derivative of the Fermi function is shown as a red line. In blue we show the convolution of the density of states chosen for the lowest temperatures in Fig.\,\ref{FigZeroField}(b) with the Fermi function at a temperature of 0.5 K. We see that the density of states has a temperature dependence that is well captured by the deconvolution procedure leading to Fig.\,\ref{FigZeroField}(b). (b) The tunneling conductance obtained at 0.1 K is shown as black discs. The black line is the convolution of the red line (which is $N_{Superc}(E)$ for 0.1 K in Fig.\,\ref{FigZeroField}(b)) with the Fermi function at a temperature of 0.1 K. For comparison, we provide in blue the density of states following Dynes inelastic scattering expression $N_{Dynes}(E)$ that best follows $N_{Superc}(E)$, with $\Gamma=$0.05 meV and $\Delta=0.16$ meV.
}
\label{FigTunnelTemp}
\end{figure}

The BCS shape of $N_{BCS}(eV)$ is found until close to $T_c$ in Pb, i.e. $N_{Superc}(eV)=N_{BCS}(eV)$. In Pb there is a small but non-zero contribution to the density of states from inelastic scattering. One can use then $N_{Dynes}(E)=\Re \left(\frac{E-i\Gamma}{\sqrt{(E-i\Gamma)^2-\Delta^2}}\right)$, with $\Gamma$ being very small in Pb, $\Gamma\approx 0.01$ meV, i.e. less than 1\% of the superconducting gap value, $\Delta=1.3$ meV\cite{PhysRevLett.41.1509}. As shown in Ref.\,\cite{doi:10.1063/5.0064511}, the influence of $\Gamma$ is practically negligible in the experiment, so that $G(V)$ can be traced as a function of temperature by using $N_{BCS}(E)=\left(\frac{E}{\sqrt{E^2-\Delta^2}}\right)$ until very close to $T_c$. It is important to note that, even close to $T_c$ and in spite of the smearing induced by the Fermi function, it is still possible to disentangle details in $N_{exp}(E)$. For example, introducing a $N_{Superc}(E)$ with a finite density of states at $E=0$ at 6 K ($T_c=$ 7.2 K) or taking a different position for the quasiparticle peaks in $N_{Superc}(E)$ does not lead to the measured $G(V)$ in Ref.\,\cite{doi:10.1063/5.0064511}. This also applies to the results shown here in Au$_2$Pb. For example, we show in Fig.\,\ref{FigTunnelTemp}(a) the tunneling conductance $G(V)$ obtained at 0.5 K as black disks. The tunneling conductance obtained by using the same $N_{Superc}(E)$ as the one used at 0.1 K is shown as a blue line. Clearly, there is a temperature induced modification. The tunneling conductance obtained by a modified $N_{Superc}(E)$ (also shown in Fig.\,\ref{FigZeroField}(b) for the curve at 0.5 K) is shown as a red line.

Other than the density of states $N(E)$ proposed in Fig.\,\ref{Schematics}(c), we can also analyze further possibilities. We show in Fig.\,\ref{FigTunnelTemp}(b) the tunneling conductance obtained in the experiment at 0.1 K as black disks. The $N_{Superc}(E)$ used to obtain the black line through convolution with the Fermi function at 0.1 K is shown as a red line. We see a sub-linear dispersion to very low energies and a close to linear increase above about 0.1 mV. We also see that $N_{Dynes}(E)$ (blue curve in Fig.\,\ref{FigTunnelTemp}(b)) leads to a flat energy dependence at low energies which does not reproduce what we find in the experiment. The $N_{Superc}(E)$ curve we find is also compatible with a d-wave order parameter, which provides a density of states that is very similar to the one shown as a red line in Fig.\,\ref{FigTunnelTemp}(b), provided that there are randomly distributed pair breaking states. Nevertheless, macroscopic experiments show an opened superconducting gap, there is no evidence for d-wave superconductivity in the bulk.

\section*{Tunneling conductance in a magnetic field}

\begin{figure}[htbp]
\includegraphics[width = \columnwidth]{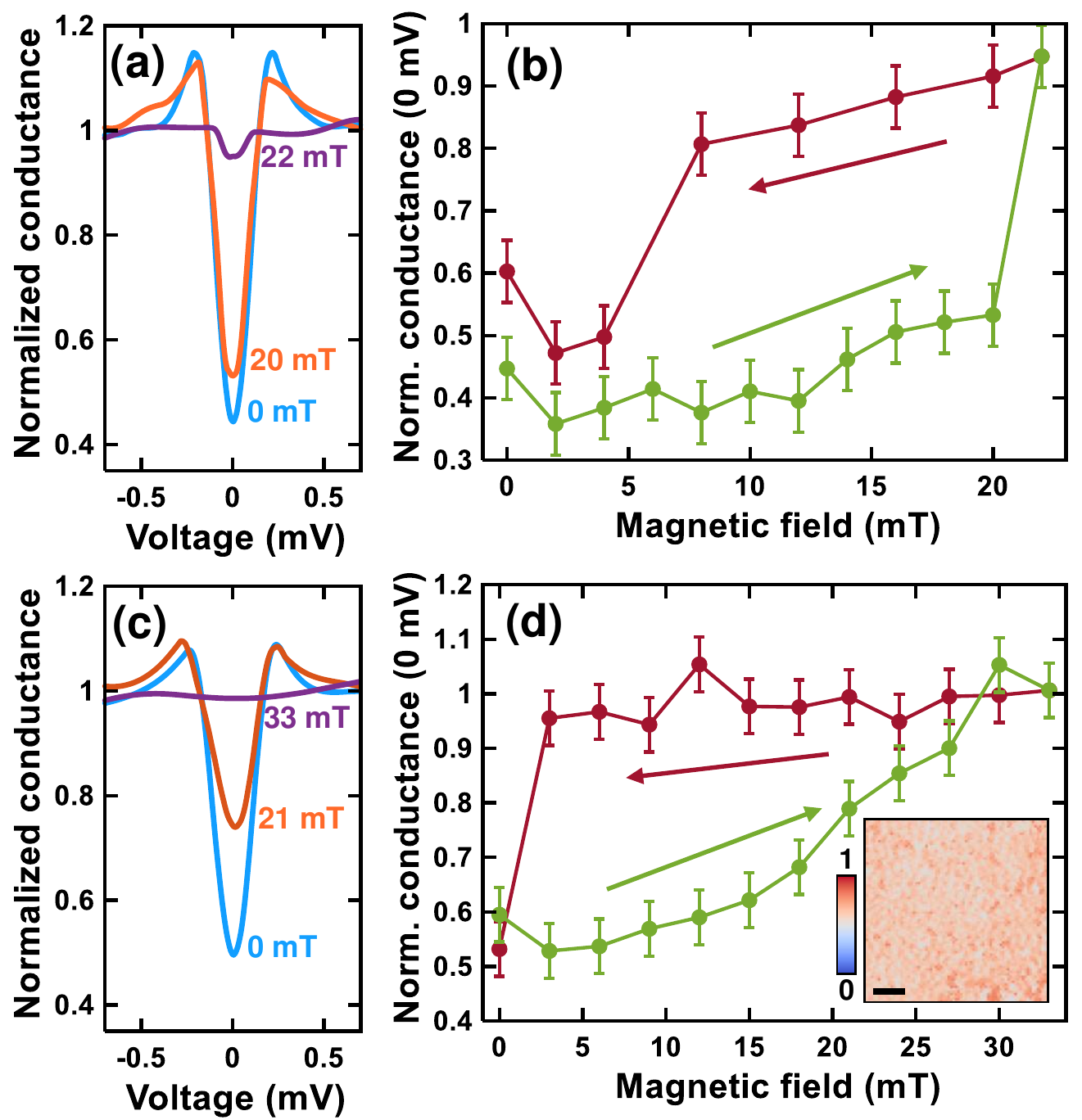}
\caption{(a) Magnetic field dependence of the superconducting tunneling conductance, with the values of the magnetic field indicated. (b) Zero bias conductance as a function of the magnetic field for increasing (green) and decreasing (dark red) magnetic fields. (c,d) Same as (a,b) in a different field of view. The inset in (d) shows a tunneling conductance map at zero bias voltage at 100 mK and at a (decreasing) magnetic field of 0.01 T. Horizontal scale bar (black line) is 50 nm long.
}
\label{FigField}
\end{figure}

To further study the superconducting properties we measure the tunneling spectroscopy in a magnetic field. We observe that superconductivity disappears most often between 20 mT and 30 mT (Fig.\,\ref{FigField}), depending on the field of view. We could find no indications for vortices or of separated normal and superconducting areas within the fields of view we investigated under magnetic fields. However, we observe a clear hysteresis when measuring the magnetic field dependence of the tunneling conductance at a fixed position (Fig.\,\ref{FigField}). Previous bulk measurements have found a first critical field of about 30 mT. There are reports about the presence of a second critical field in the range between 20 mT and 80 mT \cite{Schoop2015,Xing2016,Yu2016,Wu2018}. From these data we can estimate a coherence length of at least $\xi\approx 60$ nm.

There is clearly Meissner field expulsion until several tens of mT at low temperatures \cite{Xing2016}. Thus, we can expect strong demagnetizing effects, similar to those that can be found in elemental Pb or in Nb at small magnetic fields \cite{Prozorov2008,PhysRevLett.102.136408}. Our sample is nearly plate-like, with a demagnetizing factor close to one. This implies that the magnetic field enters the sample well below the first critical field and that we can expect an inhomogeneous field distribution and hysteretic behavior when increasing and decreasing the magnetic field due to trapped flux \cite{RevModPhys.91.015004,BRANDT349,PhysRevB.3.2231}. The large value of the coherence length suggests that the separation of the sample into normal and superconducting areas, characteristic of the intermediate state \cite{Prozorov2008,PhysRevLett.102.136408,garciacampos2020visualization}, occurs over length scales that are much larger than our fields of view. The hysteresis shows however that such well separated normal and superconducting areas are present in our sample. We also notice that small angle neutron scattering experiments did not lead to the observation of a vortex lattice \cite{Emma22}.

Clearly, Au$_2$Pb is a type I or weakly type II superconductor. This adds Au$_2$Pb to the rapidly increasing list of non-elemental materials that are type I or weak type II superconductors. We can find reports indicating such a behavior in Al$_6$Re, YNiSi$_3$, LuNiSi$_3$, SnAs, RuB$_2$, KBi$_2$, PdTe$_2$, CaBi$_2$, $\beta$-IrSn$_4$, RPd$_2$Si$_2$ ($R=Y, La, Lu$), LaRh$_2$Si$_2$, TaSe$_2$, Ag$_5$Pb$_2$O$_6$, LaRhSi$_3$, ScGa$_3$, LuGa$_3$, YbSb$_2$, BeAu, LiBi, potassium graphite intercalation compounds or SrRu$_2$O$_4$ (the latter two along a given field direction) \cite{PhysRevB.99.144519,PhysRevB.99.224505,PhysRevB.100.184514,PhysRevB.97.054506,Sun_2016,Chen2018,PhysRevB.96.220506,PhysRevB.94.144506,C6CP02856J,Tran_2013,doi:10.1021/acs.chemmater.0c00179,doi:10.1143/JPSJ.50.3245,PhysRevB.34.4566,PhysRevB.45.4803,PhysRevB.72.180504,PhysRevB.83.064522,PhysRevB.85.174514,doi:10.1143/JPSJ.56.419,PhysRevB.99.134509,TakashiAkima1999,garciacampos2020visualization}.


%

\end{document}